\documentclass[twocolumn,twoside,slac_two]{revtex4}
\usepackage{graphicx}
\usepackage{fancyhdr}
\usepackage{subfig}
\begin{document}

\title{ LHCb status and charm physics program }

\author{Patrick Spradlin (on behalf of the LHCb collaboration)}
\affiliation{Oxford University, Oxford, United Kingdom}

\begin{abstract}
  LHCb is a dedicated flavor physics experiment that will observe the
  $14\,\mathrm{TeV}$ proton--proton collisions at CERN's Large Hadron Collider
  (LHC).  Construction of the
  LHCb detector is near completion, commissioning of the detector is well
  underway, and LHCb will be fully operational and ready to take data in
  advance of the projected May 2008 turn-on date for the LHC.
  The LHCb software trigger will feature a dedicated channel for events
  containing $D^{\ast}$ mesons that will dramatically enhance the
  statistical reach of LHCb in many charm physics measurements.
  The LHCb charm physics program is initially focused on mixing and CP
  violation measurements in two body decay modes of $D^0$.  A much broader
  program is possible and will be explored as manpower allows.
  We intend to use both promptly produced charm and secondary charm from $B$
  meson decays in measurements.  Initial studies have focused on using
  secondary $D^{\ast+}$ mesons for mixing measurements in two body decays.
  Preliminary Monte Carlo studies indicate that LHCb may obtain
  a statistical precision of
  \mbox{$\sigma_{\mathrm{stat}}({x'}^2) = \pm 0.064 \times 10^{-3}$} and
  \mbox{$\sigma_{\mathrm{stat}}(y') = \pm 0.87 \times 10^{-3}$}
  from a time dependent mixing analysis of wrong sign two body
  \mbox{$D^0 \rightarrow \pi^- K^+$} decays
  and a statistical precision of
  \mbox{$\sigma_{\mathrm{stat}}(y_{\mathrm{CP}}) = \pm 0.5 \times 10^{-3}$}
  from a ratio of the lifetimes of $D^0$ decays to the final states $K^- K^+$
  and $K^- \pi^+$ in $10\,\mathrm{fb}^{-1}$ of data.
\end{abstract}

\maketitle

\thispagestyle{fancy}

\section{ LHCb status \label{sec:status} }

\begin{figure}

  \includegraphics[width=80mm]{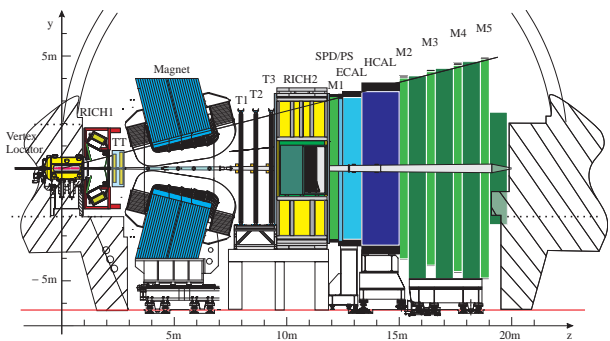}

\caption[ LHCb detector layout ]{
  LHCb detector layout, showing all of the detector components.
  Also shown are the direction of the $y$ and $z$ coordinate axes; the $x$ axis
  completes the right handed coordinate system~\cite{LHCbTDR:2003ro}.
 \label{fig:status:det}}
\end{figure}

As the dedicated flavor experiment at CERN's Large Hadron Collider (LHC),
LHCb is designed to optimally
exploit the large $b\overline{b}$ production cross--section in the LHC
\mbox{$14\,\mathrm{TeV}$} proton--proton collisions for precision measurements
of $b$ hadron properties.
Figure~\ref{fig:status:det} shows the layout of the LHCb detector.  In high
energy hadronic collisions that produce $b\overline{b}$ pairs, the $b$-- and
$\overline{b}$--hadrons are predominantly produced into the same forward
cone---a fact that led to LHCb's single--arm spectrometer
design~\cite{LHCbTDR:1998tp}.
The angular acceptance of the detector
extends from approximately \mbox{$10\,\mathrm{mrad}$} around
the beam axis to \mbox{$300\,\mathrm{mrad}$} in the magnetic bending plane
and to \mbox{$250\,\mathrm{mrad}$} in the non--bending plane.

Many of the features that make LHCb an excellent $B$ physics laboratory also
make LHCb well--suited for many charm physics studies at unprecedented levels
of precision.  The silicon Vertex Locator (VELO) will provide the excellent
vertex resolutions necessary for time dependent measurements---an estimated 
\mbox{$45\,\mathrm{fs}$} proper time resolution for
\mbox{$D^0 \rightarrow K^- \pi^+$} decays where the $D^0$ mesons are produced
in $b$-hadron decays.
The LHCb tracking system will
supply precise momentum measurements---an estimated \mbox{$6\,\mathrm{MeV}$}
mass resolution for two body decays of $D^0$ mesons.  The LHCb Ring Imaging
Cherenkov (RICH) detectors will
provide excellent $K$--$\pi$ discrimination over a wide momentum range
from \mbox{$2\,\mathrm{GeV/c}$} to \mbox{$100\,\mathrm{GeV/c}$}.  Finally, the
LHCb trigger system will have a high statistics charm stream,
described in Section~\ref{sec:trig}, so that the large charm production in
LHC collisions can be exploited for precision measurements.

As of September 2007, construction of the LHCb detector is well advanced with
commissioning activities underway for all sub--detectors.  LHCb is on--schedule
to be complete and ready for data taking by the projected LHC turn-on date in
May 2008.

\section{ LHCb trigger and $D^{\ast}$ stream \label{sec:trig} }

LHCb will have a two stage trigger:  a fast hardware trigger called
the Level 0 Trigger (L0) followed by a software High Level Trigger (HLT).
Although the triggers are designed to favor $b\overline{b}$ events, the HLT
will feature a dedicated $D^{\ast}$ stream for selecting charm events at a
high rate.

At design operation, LCHb will observe bunch crossings at
\mbox{$40\,\mathrm{MHz}$} with a luminosity of
\mbox{$2 \times 10^{32}\,\mathrm{cm}^{-2}\,\mathrm{s}^{-1}$}.  The L0 trigger
is designed to reduce the \mbox{$40\,\mathrm{MHz}$} input rate to
approximately \mbox{$1\,\mathrm{MHz}$} while efficiently favoring
$b\overline{b}$ events.  Using the fact that the decay products of $b$-hadrons
typically have significant transverse momentum, the L0 trigger reads data
from the calorimeters and the muon detectors to identify quickly individual
candidate hadrons, electrons, photons, and muons that have a few
$\mathrm{GeV}$ of transverse energy or momentum.  The L0 $E_T$ thresholds must
be tuned on real data, but, to illustrate the expected scale, current detector
studies indicate a trigger threshold for hadrons of
\mbox{$E_T > 3.5\,\mathrm{GeV}$}.

The HLT software trigger performs an event reconstruction to identify events
of specific physics interest, reducing the \mbox{$1\,\mathrm{MHz}$} L0
output/HLT input rate to approximately \mbox{$2\,\mathrm{kHz}$}.
The HLT has access to all of the detector information and uses it to
reconstruct final state particle candidates and the locations of the primary
proton-proton interactions.  These objects are then used in several
parallel channels to identify events of specific physical interest.
Although the configuration of the HLT channels has yet to be finalized, the
HLT will contain a high yield stream for charm events.
In the current configuration, \mbox{$300\,\mathrm{Hz}$}, $15\%$ of
the recorded \mbox{$2\,\mathrm{kHz}$} of LHCb events, are allocated to an
inclusive $D^{\ast}$ stream.
Table~\ref{tab:trig:yld} shows potential yields of some key charm decays in
a preliminary configuration the $D^{\ast}$ stream.  These estimates are based
on the performance of the HLT on fully simulated LHCb events.
LHCb will record at least
\mbox{$50 \times 10^6$} fully reconstructed
\mbox{$D^{\ast+} \rightarrow \pi_{\mathrm{s}}^{+} D^0(h^-{h'}^+)$}
\cite{CPConj:note} candidates
per $2\,\mathrm{fb}^{-1}$ (a nominal year of LHCb data at design luminosity),
where the $D^{\ast+}$ originates in a $b$-hadron decay and
\mbox{$h, h' \in \{K, \pi\}$}.  The subscript on
the $\pi_{\mathrm{s}}^{+}$ labels it as the tagging `slow' pion.
Studies indicate that this HLT configuration also
yields a similar number of reconstructed prompt $D^{\ast+}$ candidates in each
mode.

\begin{table}
\caption[ HLT yield of secondary charm events ] {
  Estimated yield of reconstructed secondary
  \mbox{$D^{\ast+} \rightarrow \pi_{\mathrm{s}}^{+} D^0(h^-{h'}^+)$} candidates
  in $2\,\mathrm{fb}^{-1}$ of LHCb data passing the LHCb trigger sequence.
\label{tab:trig:yld}}

  \begin{tabular*}{\linewidth}{@{\extracolsep{\fill}}l|r} \hline\hline
    Two body $D^0$ mode &
    HLT Yield in $2\,\mathrm{fb}^{-1}$ \\ \hline

    $D^0 \rightarrow K^- \pi^+$ &
    $50 \times 10^6$ \\

    $D^0 \rightarrow K^- K^+$ &
    $5 \times 10^6$ \\

    $D^0 \rightarrow \pi^-\pi^+$ &
    $2 \times 10^6$ \\

    $D^0 \rightarrow \pi^-K^+$ &
    $0.2 \times 10^6$ \\ \hline\hline
  \end{tabular*}

\end{table}

\section{ $D^{\ast+}$ event selection \label{sec:sel} }

The charm physics program at LHCb is initially focused on mixing and CP
violation measurements in two body decays of $D^0$.
LHCb note~\cite{Spradlin:1045412} details a preliminary selection of wrong
sign (WS) \mbox{$D^0 \rightarrow \pi^- K^+$} decays from
\mbox{$D^{\ast+} \rightarrow D^0 \pi_{\mathrm{s}}^+$}, where the $D^{\ast+}$
originates from a $B$ meson decay.
The intent of the selection is to provide a sample of candidates suitable for
a time dependent mixing analysis (see Section~\ref{sec:mix:ws}).
The performance of the selection on fully simulated LHCb data predicts a
yield of approximately $230,000$ true WS decays in
\mbox{$10\,\mathrm{fb}^{-1}$} of LHCb data (5 years of nominal LHCb data
taking) with a background-to-signal ratio of
\mbox{$1.07 < B / S < 5.28$} at the $90\%$ confidence level.

  \subsection{ Selection backgrounds \label{sec:sel:bkg} }

    The primary backgrounds accepted by this selection result from combinatoric
    coincidences and can be divided into two classes:  random slow pion
    backgrounds where properly reconstructed right sign 
    \mbox{$D^0 \rightarrow K^- \pi^+$} decays are combined with random pions
    produced somewhere else in the event to mimic a $D^{\ast\pm}$ decay, and
    pure combinatoric coincidences where the candidate $D^0$ decay products
    come from different decays in the event.  These
    backgrounds will be separated from the signal with the typical method of
    fitting the reconstructed $D^0$ mass ($m_{D^0}$) and $D^{\ast+}$-$D^0$
    mass difference ($\Delta m$) distributions.

  \subsection{ $D^{\ast+}$ decay vertex \label{sec:sel:vtx} }

    In order to perform a time dependent mixing measurement, both the creation
    and decay vertices of the $D^0$ must be precisely determined.  The $D^0$
    decay vertex can be measured from its decay products very precisely in the
    VELO.
    Table~\ref{tab:mix:vtx:res} shows the results of vertex resolution
    studies in fully simulated LHCb events.
    The coordinate system in the table is that defined in
    Figure~\protect\ref{fig:status:det}, with the primary proton-proton
    collisions along the $z$ axis.
    The $D^0$ decay vertex can
    be determined with a resolution of {$257\ \mu\mathrm{m}$} along the
    beam axis.
    To provide some context for this value, the mean laboratory flight distance
    for a $60\,\mathrm{GeV/c}$ $D^0$ (the mean momentum of $D^0$ from
    $D^{\ast+}$ from $B$ mesons) is approximately \mbox{$4\ \mathrm{mm}$}.

    In contrast, the $D^{\ast+}$ decay vertex ($D^0$ creation vertex) is
    poorly determined from its decay products.  The small mass difference
    between the $D^{\ast+}$ and its decay products leads to a narrow laboratory
    frame angle between the $D^0$ and $\pi_{\mathrm{s}}^0$ momenta.
    The $D^{\ast+}$ column in Table~\ref{tab:mix:vtx:res} shows that the
    resolution of the $D^{\ast+}$ decay vertex estimated only from its decay 
    products is {$4232\ \mu\mathrm{m}$}, the same size as the mean 
    laboratory flight distance of $D^0$ mesons.  The precision of the
    $D^{\ast+}$ vertex must be improved by including additional tracks from
    particles created with the $D^{\ast+}$.  For $D^{\ast+}$ from $B$ decays,
    this means finding additional charged particles created at the $B$ decay
    vertex.

\begin{figure}
    \begin{center}
      \subfloat[]{\label{fig:mix:vtx:lt:lt}\includegraphics[width=80mm]{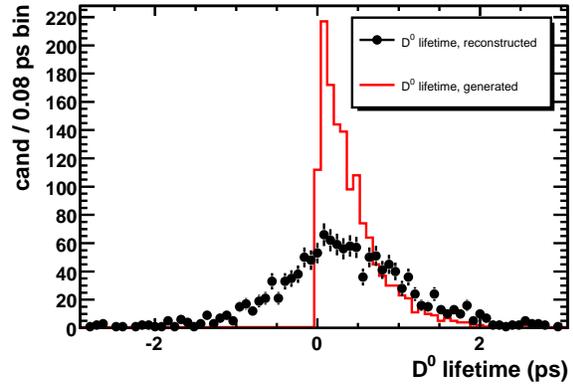}}

      \subfloat[]{\label{fig:mix:vtx:lt:impLT}\includegraphics[width=80mm]{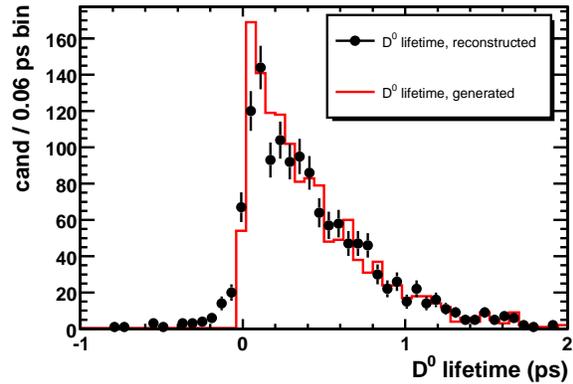}}
    \end{center}

    \caption{
	Distributions of the proper times for simulated $D^0$ mesons
	from \mbox{$B \rightarrow D^{\ast+} X$} decays.
	In each plot, the solid lines are the generated proper times and
	the points are the estimated $D^0$ proper times
	using~\protect\subref{fig:mix:vtx:lt:lt} the poorly estimated
	$D^{\ast+}$ decay vertex, or~\protect\subref{fig:mix:vtx:lt:impLT}
	the precisely estimated parent $B$ decay vertex as the $D^0$
	production vertex.
     \label{fig:mix:vtx:lt}}
    \end{figure}

    In studies of fully simulated events, $63\%$ of
    \mbox{$B \rightarrow D^{\ast+} X$} decays in triggered events produce an
    additional reconstructed charged particle at the $B$ meson decay vertex
    that can be used to improve the precision of the estimated $D^0$ birth
    vertex.
    As shown in the $B_{\mathrm{part}}$ column of Table~\ref{tab:mix:vtx:res},
    using such additional tracks dramatically improves the precision of the
    estimated $D^0$ production vertex, and, consequently, the measured $D^0$
    proper time.
    The subscript on $B_{\mathrm{part}}$ signifies that the
    parent $B$ is partially reconstructed.
    Figure~\ref{fig:mix:vtx:lt} shows the dramatic improvement in measured
    proper time obtained by using the $B_{\mathrm{part}}$ decay vertex as the
    $D^0$ production vertex.
    In these plots, the reconstructed proper time is signed to represent
    whether the $D^{0}$ momentum and flight direction are aligned (positive
    proper time) or anti-aligned (negative proper time).
    When the $D^{\ast+}$ decay vertex is used in calculating the $D^0$ proper
    time, its resolution dominates the exponential decay distribution as
    shown in Figure~\ref{fig:mix:vtx:lt:lt}.
    When the $B_{\mathrm{part}}$ decay vertex is used, as in
    Figure~\ref{fig:mix:vtx:lt:impLT}, the proper time resolution is relatively
    narrow and the reconstructed proper time distribution closely reproduces
    the generated proper time.
    Preliminary work in~\cite{Spradlin:1045412} has demonstrated that this
    partial $B$ reconstruction can be done with an $80\%$ efficiency, and that
    LHCb can use $D^{\ast+}$ mesons from $B$ decays for time dependent
    analyses.

    \begin{table}
    \caption[ Resolutions of $D^0$, $D^{\ast+}$, and $B_{\mathrm{part}}$
      vertices, and of $D^0$ proper time ]{
	Estimated resolutions of $D^0$, $D^{\ast+}$, and $B_{\mathrm{part}}$
	vertices, and of $D^0$ proper time in simulated LHCb data.
     \label{tab:mix:vtx:res}}

      \begin{tabular*}{\linewidth}{@{\extracolsep{\fill}}r|rr|r} \hline\hline

        \hspace{3em} &
        $D^{0}$ &
        $D^{*+}$ &
        $B_{\mathrm{part}}$ \\ \hline

        $x$ &
        $21.6\ \mu\mathrm{m}$ &
        $187.\ \mu\mathrm{m}$ &
        $18.1\ \mu\mathrm{m}$ \\

        $y$ &
        $16.9\ \mu\mathrm{m}$ &
        $144.\ \mu\mathrm{m}$ &
        $18.4\ \mu\mathrm{m}$ \\

        {$z$} &
        {$257.\ \mu\mathrm{m}$} &
        {$4232.\ \mu\mathrm{m}$} &
        {$237.\ \mu\mathrm{m}$} \\

        {$\tau$} &
        \multicolumn{2}{c|}{{$0.465\ \mathrm{ps}$}} &
        {$0.045\,\mathrm{ps}$} \\ \hline\hline
      \end{tabular*}

    \end{table}

\section{ Charm mixing measurements at LHCb \label{sec:mix} }

Mixing observables are commonly expressed in terms of two dimensionless
parameters:  the mass difference parameter, $x$, and the full width difference
parameter, $y$, defined by
\[
  x = \frac{2(m_1 - m_2)}{\Gamma_1 + \Gamma_2}, \hspace{2em} 
  y = \frac{\Gamma_1 - \Gamma_2}{\Gamma_1 + \Gamma_2},
\]
where the subscripts denote the mass eigenstates of the
\mbox{$D^0$-$\overline{D^0}$} system.  Various measurements are sensitive to
different combinations of these variables, and each of these should
be explored with the highest possible precision to gain a full understanding of
the charm mixing phenomenon.
Preliminary work at LHCb has focused on
the measurement of mixing parameters in a time dependent analysis of two body
WS decays (Section~\ref{sec:mix:ws} below) and in an analysis of
the ratios of two body lifetimes (Section~\ref{sec:mix:ltrat}
below).  However, future plans include multi-body mixing measurements
(Section~\ref{sec:multi}).

  \subsection{ Time dependent WS $D^0 \rightarrow \pi^- K^+$  \label{sec:mix:ws} }

    The time dependent analysis of WS \mbox{$D^0 \rightarrow \pi^- K^+$} 
    decays is one of the long established methods
    of searching for $D^0$-$\overline{D^0}$ mixing~\cite{Aitala:1996fg,Godang:1999yd,Link:2004vk,Zhang:2006dp,Aubert:2007wf}.
    If $D^0$ and $\overline{D^0}$ mix, a meson created as a $D^0$ may decay to
    the WS final state \mbox{$\pi^- K^+$} either directly, by a doubly Cabibbo
    suppressed (DCS) decay, or indirectly, by mixing into a $\overline{D^0}$
    meson that undergoes a Cabibbo favored (CF) decay.
    Interference between the two processes leads to a decay time
    dependence that can be expanded to leading order in the small parameters
    $x$ and $y$ (in the absence of CP violation) as
    \[
      r_{\mathrm{WS}}(t) \propto \mathrm{e}^{-\Gamma t} \left(R_D 
      + \sqrt{R_D} y' (\Gamma t)
      + \frac{1}{2} R_M (\Gamma t)^2 \right),
    \]
    where $R_D$ is the ratio of the DCS decay rate to the CF decay rate,
    \mbox{$R_M = (x^2 + y^2) / 2$} \mbox{$= ({x'}^2 + {y'}^2) / 2$} is the
    mixing rate, and $x'$ and $y'$ are rotated with respect to the parameters
    $x$ and $y$ by the relative strong phase between the CF and DCS decays,
    $\delta$:
    \begin{eqnarray*}
      x' & \equiv & x \cos\delta + y \sin\delta, \\
      y' & \equiv & y \cos\delta - x \sin\delta.
    \end{eqnarray*}
    Hence, a detailed analysis of the WS $D^0$ proper time distribution is
    sensitive to ${x'}^2$ and $y'$.  The strong phase $\delta$ that relates
    these quantities to the mixing parameters $x$ and $y$ must be measured
    independently.  Since ${x'}^2$ enters the decay time distribution
    only in \mbox{$R_M = ({x'}^2 + {y'}^2) / 2$}, the values of ${x'}^2$
    measured by this method are highly anti-correlated to the values of $y'$.

\begin{figure}
    \begin{center}
      \includegraphics[width=80mm]{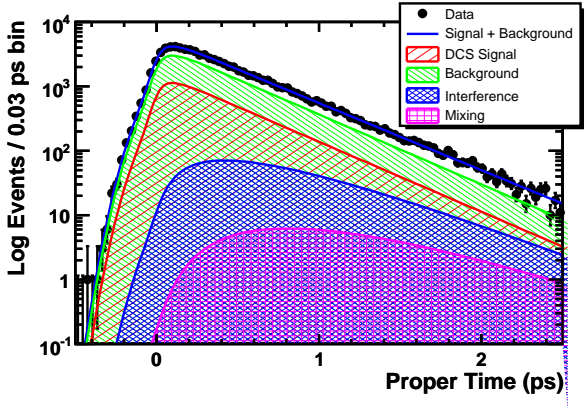}
    \end{center}

    \caption[ WS Toy Monte Carlo sample ] {
      An example of a Toy Monte Carlo sample from
      the time dependent WS analysis study.
     \label{fig:mix:ws:extoy}}
    \end{figure}

    Because of the small values of $x$ and $y$, it is only recently
    that this method has yielded values over $3\sigma$ from
    \mbox{${x'}^2 = 0$}, \mbox{$y' = 0$}~\cite{Aubert:2007wf}.
    The large charm statistics at LHCb should be able to improve significantly
    this picture.
    The selection described in Section~\ref{sec:sel} estimates that
    a time dependent WS mixing analysis on \mbox{$10\,\mathrm{fb}^{-1}$} of
    LHCb data would incorporate approximately $230,000$ $D^{\ast+}$ tagged
    WS decays from $B$ decays.  The \mbox{$10\,\mathrm{fb}^{-1}$} signal and
    background yields, the proper
    time resolution, and the proper time acceptance of this selection were
    used in a toy Monte Carlo study to estimate the LHCb statistical
    sensitivity to ${x'}^2$ and $y'$:
    \begin{eqnarray*}
      \sigma_{\mathrm{stat}}({x'}^2) = \pm 0.064 \times 10^{-3}, \\
      \sigma_{\mathrm{stat}}(y') = \pm 0.87 \times 10^{-3}.
    \end{eqnarray*}
    The toy study verifies the expected large negative correlation between
    $x'^2$ and $y'$: \mbox{$\mathrm{Corr}(x'^2, y') = -0.95$}.
    Figure~\ref{fig:mix:ws:extoy} shows a toy sample from this
    study~\cite{Spradlin:1045412}.

  \subsection{ Lifetime ratio \label{sec:mix:ltrat} }

    Just as it does in the $K^0$ and $B^0$ systems, the presence of mixing in
    the $D^0$ system can lead to different lifetimes for states of different
    CP content.  These differences are related to the mixing parameters $x$
    and $y$.  In two body $D^0$ decays, the ratio of the lifetime of
    decays to the CP-even eigenstate $K^- K^+$, $\tau(K^- K^+)$, and the
    lifetime of decays to the non-CP CF eigenstate $K^- \pi^+$,
    $\tau(K^- \pi^+)$, form the observable
    \[
      y_{\mathrm{CP}} \equiv \frac{\tau(K^- \pi^+)}{\tau(K^- K^+)} - 1,
    \]
    which is related to the mixing parameters by
    \[
      y_{\mathrm{CP}} = y \cos\phi - \frac{1}{2} A_M x \sin\phi,
    \]
    where $\phi$ is a weak phase and $A_M$ parameterizes CP violation in
    $D^0$ mixing.
    Decays to the CP-even eigenstate $\pi^- \pi^+$ may also be used in place of
    $K^- K^+$ for an independent measurement of $y_{\mathrm{CP}}$.
    The CP violating parameters $A_M$ and $\phi$
    must be measured independently, but in the no CP violation limit,
    \mbox{$A_M = 0$} and \mbox{$\phi = 0$}, and the lifetime ratio is a direct
    measurement of \mbox{$y_{\mathrm{CP}} = y$}.

    This two body lifetime ratio measurement has been carried out at several
    previous experiments \cite{Aitala:1999dt,Link:2000cu,Csorna:2001ww,Aubert:2003pz}
    culminating in Belle's recent $3.2\sigma$ measurement of $y_{\mathrm{CP}}$
    in \cite{Staric:2007dt}.  LHCb expects to use its statistical advantage
    over current experiments to improve the precision of the lifetime
    ratio measurement of $y_{\mathrm{CP}}$.

    The LHCb statistical sensitivity to $y_{\mathrm{CP}}$ has been estimated
    with a toy Monte Carlo study similar to the the WS mixing toy study
    described in Section~\ref{sec:mix:ws}.  The selection described in
    Section~\ref{sec:sel}, with appropriate modifications of the final state
    particle identification criteria, yields approximately
    \mbox{$8 \times 10^6$} $D^{\ast+}$ tagged \mbox{$D^0 \rightarrow K^- K^+$}
    decays and \mbox{$3 \times 10^6$} $D^{\ast+}$ tagged
    \mbox{$D^0 \rightarrow \pi^- \pi^+$} decays
    originating from $b$-hadron decays in $10\,\mathrm{fb}^{-1}$ of LHCb data.
    The signal to background ratios of the selection are \mbox{$S / B = 4.8$}
    for the $K^-K^+$ mode and \mbox{$S / B = 2.6$} for the $\pi^-\pi^+$ mode.
    Using the estimated \mbox{$D^0 \rightarrow K^- K^+$} yield and signal to
    background ratio with proper time resolution and acceptance functions
    determined from fully simulated LHCb events,
    the estimated statistical precision of $y_{\mathrm{CP}}$ is
    \mbox{$\sigma_{\mathrm{stat}}(y_{\mathrm{CP}}) = \pm 0.5 \time 10^{-3}$}.

\section{ Searches for CP violation at LHCb \label{sec:cpv} }

The Standard Model (SM) predicts any CP violation in charm interactions to
be very small.
Observable CP violation at the level of $1\%$ would be an unambiguous sign
of new physics~\cite{Grossman:2006jg}.
Each of the mixing measurements performed at LHCb will be analyzed in charge
conjugate subsets to measure possible CP violating effects.
In addition, LHCb will perform time integrated CP violation searches in as
many charm decays as are possible.
Initial studies have focused on searching for CP violation in two body
decays of $D^{\ast+}$ tagged $D^0$ mesons, in particular the CP eigenstate
decays \mbox{$D^0 \rightarrow K^- K^+$} and
\mbox{$D^0 \rightarrow \pi^- \pi^+$}.
These singly Cabibbo suppressed decays, in which a small CP violation is
predicted by the SM, are particularly sensitive to CP violation
enhanced by well-motivated new physics scenarios~\cite{Grossman:2006jg}.
Experimental measurements in this channel have steadily reduced the
upper limit of CP violation with increasing data set sizes and improved
treatments of systematic uncertainties
\cite{Aitala:1997ff,Link:2000aw,Csorna:2001ww,Acosta:2004ts,Aubert:2007if}.
However, CP violation at the order of $1\%$ has not been ruled out.

With 
\mbox{$8 \times 10^6$} tagged \mbox{$D^0 \rightarrow K^- K^+$} decays and
\mbox{$3 \times 10^6$} tagged \mbox{$D^0 \rightarrow \pi^- \pi^+$} decays
(Section~\ref{sec:mix:ltrat}),
LHCb will have the
statistical power to search for CP asymmetries to order $\mathcal{O}(0.0004)$
or below, provided the systematic uncertainties can be controlled to this
level.

Charm meson production asymmetries and final state particle detector
asymmetries, particularly the detector asymmetries associated with the tagging
slow pion, are expected to be the primary sources of systematic uncertainties
in CP asymmetry measurements.  Methods of measuring the production and
detections asymmetries precisely from data are under development.
Also, advantage may be gained by comparing the asymmetries of related decays.
For example, the decays \mbox{$D^0 \rightarrow K^- K^+$} and
\mbox{$D^0 \rightarrow \pi^- \pi^+$} are subject to the same production
and slow pion detection asymmetries, so the difference of their
measured asymmetries will have a much smaller systematic uncertainty than
either asymmetry measured separately.  Although this difference will be small,
it is an observable that can be measured very precisely, and, if found to be
significantly different from zero, can provide evidence of direct CP violation
in at least one of the two decay channels.

\section{ Multi-body channels \label{sec:multi} }

LHCb will also investigate the use of $D$ meson decays to three or more
final state hadrons in mixing and CP violation measurements.
However, development of multi-hadron charm analyses is less advanced
than the two body $D^0$ program.
For example, a time dependent amplitude analysis of the three body decay
\mbox{$D^0 \rightarrow K_S^0 \pi^+ \pi^-$} is directly sensitive to the
mixing parameters $x$ and $y$.
This decay mode should be efficiently reconstructible at LHCb.
Preliminary work is also under way to develop selections for $D^0$ decays to
four hadrons, both for HLT triggering and for analysis.
Further development in four body decays will investigate the feasibility of
time dependent amplitude analyses for mixing measurements.  The technology of
four body amplitude analyses is already quite
advanced~\cite{Rademacker:2006zx}.

In four body decays, CP violation searches will include analyses
of quantities that are odd under the time reversal operation in addition to
complete amplitude analyses of the decays.
Although studies are still in their earliest stages, LHCb should be
able to reconstruct with acceptable signal-to-background ratios the
decays \mbox{$D^0 \rightarrow K_S^0 h^+ {h'}^-$}
and three body decays of charged $D^+$ containing at least one kaon.
Amplitude analyses of these modes will expand the scope
for CP violation searches in charm decays.
\section{ Conclusions \label{sec:outlook} }

\begin{figure}
\begin{center}
  \includegraphics[width=80mm]{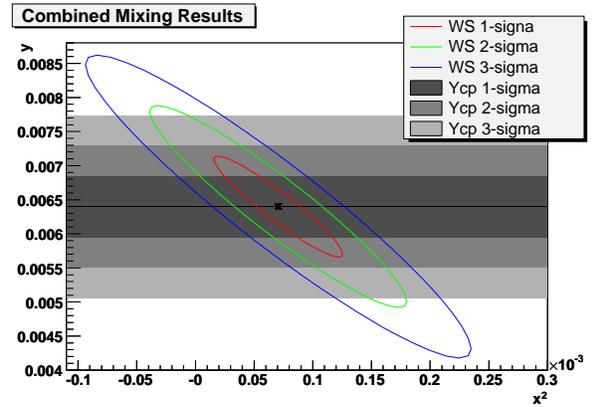}
\end{center}

\caption{
  Contours representing the $1\sigma$, $2\sigma$, and $3\sigma$ regions for
  the toy WS mixing study (ellipses)
  and for the toy lifetime ratio study (bands).
  The central values are \mbox{$\sqrt{{x}^2} = 8.4 \times 10^{-3}$} and
  \mbox{${y} = 6.4 \times 10^{-3}$}, the very preliminary averages
  of~\protect\cite{Asnerf:Felhc}.
  For purposes of this plot, it is assumed that \mbox{${x'}^2 = x^2$} and
  \mbox{$y' = y_{\mathrm{CP}} = y$}.
 \label{fig:outlook:contour}}
\end{figure}

The LHCb trigger will provide LHCb with charm physics data sets of
unprecedented statistics.
Current analysis in charm physics has focused on mixing measurements and
CP violation searches in two body decays \mbox{$D^0 \rightarrow h^- {h'}^+$},
but a broader charm physics program is envisaged.
Toy Monte Carlo studies indicate that with
\mbox{$10\,\mathrm{fb}^{-1}$} LHCb can achieve a statistical precision of
\mbox{$\sigma_{\mathrm{stat}}({x'}^2) = \pm 0.064 \times 10^{-3}$} and
\mbox{$\sigma_{\mathrm{stat}}(y') = \pm 0.87 \times 10^{-3}$} with a two
body wrong sign mixing analysis, and a statistical
precision of
\mbox{$\sigma_{\mathrm{stat}}(y_{\mathrm{CP}}) = \pm 0.5 \times 10^{-3}$} with
a two body lifetime ratio measurement.
Figure~\ref{fig:outlook:contour}
summarizes these precisions by showing the intersection of the
$1\sigma$ ellipse in \mbox{$({x'}^2, y')$}, also scaled to $2\sigma$
and $3\sigma$, and the toy lifetime ratio $1\sigma$ band in $y_{\mathrm{CP}}$,
also scaled to $2\sigma$ and $3\sigma$.

\bigskip 

\begin{thebibliography}{10}
\bibitem{CPConj:note}
The use of charge conjugate modes is implied unless otherwise indicated.
\end{thebibliography}


\end{document}